\documentclass[12pt,a4paper,final]{iopart}

\usepackage{iopams}  
\usepackage{graphicx}
\usepackage[colorlinks=true,linkcolor=blue,urlcolor=blue,citecolor=blue]{hyperref}

\begin{document}

\title{Non-equilibrium steady state of a driven levitated particle with feedback cooling}

\author{Jan Gieseler}
\address{ETH Z{\"u}rich, Photonics Laboratory, 8093 Z{\"u}rich, Switzerland}
\ead{jangi@ethz.ch}

\author{Lukas Novotny}
\address{ETH Z{\"u}rich, Photonics Laboratory, 8093 Z{\"u}rich, Switzerland}
\ead{lnovotny@ethz.ch}

\author{Clemens Moritz}
\address{Faculty of Physics, University of Vienna, Boltzmanngasse 5, 1090 Vienna, Austria}
\ead{clemens.moritz@univie.ac.at}

\author{Christoph Dellago}
\address{Faculty of Physics, University of Vienna, Boltzmanngasse 5, 1090 Vienna, Austria}
\ead{christoph.dellago@univie.ac.at}

\newcommand{\eqref}[1]{(\ref{#1})}


\date{\today}

\begin{abstract}
Laser trapped nanoparticles have been recently used as model systems to study fundamental relations holding far from equilibrium. Here we study, both experimentally and theoretically, a nanoscale silica sphere levitated by a laser in a low density gas. The center of mass motion of the particle is subjected, at the same time, to feedback cooling and a parametric modulation driving the system into a non-equilibrium steady state. Based on the Langevin equation of motion of the particle, we derive an analytical expression for the energy distribution of this steady state showing that the average and variance of the energy distribution can be controlled separately by appropriate choice of the friction, cooling and modulation parameters. Energy distributions determined in computer simulations and measured in a laboratory experiment agree well with the analytical predictions. We analyse the particle motion also in terms of the quadratures and find thermal squeezing depending on the degree of detuning. 
\end{abstract}

\maketitle

\section{Introduction}

In a macroscopic system, thermodynamic quantities such as the work carried out during a thermodynamic transformation or the heat exchanged with a heat bath have well defined values due to the statistics of large numbers. For instance, if we repeatedly carry out  a certain thermodynamic transformation always starting from the same initial state and following the same protocol, the work performed on the system will always be the same. In small systems, on the other hand, thermodynamic quantities typically fluctuate. Then the work and heat of a thermodynamic transformation, carried out, for instance, by stretching a single biomolecule in solution, need to be characterised with a statistical distribution rather than a single value. Even small systems, however, are subject to the basic laws of thermodynamics and, on the average, obey the second law usually formulated in terms of inequalities. As realised by Jarzynski, more specific results can be derived for the fluctuations of work and other quantities that transform the inequalities of thermodynamics into equalities \cite{Jarzynski:PRL:1997,Jarzynski:PRE:1997,Crooks:PRE:2000,Crooks:JStatPhys:1998}, which remain valid arbitrarily far from equilibrium. Such so-called fluctuation theorems have now been derived for several quantities, such as heat, work and entropy \cite{Evans2002,Seifert2012}, shedding new light on the significance of irreversibility and the second law at the nanoscale \cite{Bustamante2005,Jarzynski:2011}. Besides their fundamental importance, fluctuation theorems also provide the basis for the interpretation of single-molecule experiments \cite{HummerSzabo:PNAS:2001,Liphardt:S:2002,Alemany2012} as well as for the development of novel non-equilibrium computer simulation methods \cite{DellagoHummer2013}.

Experimentally, fluctuation relations have been studied in a variety of systems mainly in the over-damped regime, such as a particle dragged through a liquid \cite{Sevick2002} or a biomolecule in solution \cite{Liphardt:S:2002}, where the system is strongly coupled to a thermalising environment. Recently, several experimental setups for the investigation of non-equilibrium fluctuations under low-damping conditions were proposed \cite{Ciliberto2010,NAT_NANOTECH, Lechner_2013, Dechant_2014}. Due to their weak coupling to the heat bath, such systems hold the promise to enable investigation of the statistics of non-equilibrium fluctuations in the quantum regime. Also, the precise control over the dynamics that can be achieved in such systems permits to construct situations in which microscopic reversibility does not hold. 

Here, we study, using theory, simulation and experiment a levitated nanoparticle in the low-friction regime \cite{Gieseler_PRL2012}. In particular, we derive analytical expressions for the energy and phase-space distribution of the system in non-equilibrium steady states. Based on these distributions one can relate heat, entropy and energy to each other, thereby providing additional insight into the physics underlying the fluctuation theorems. The particle, consisting of a dielectric material, oscillates in a laser trap and is surrounded by a low-density gas, which exerts frictional and random thermal forces on the particle. The amount of friction can be controlled by changing the pressure of the gas. In addition, the particle is subjected to a nonlinear feedback cooling mechanism and a parametric modulation. Together, these effects allow to bring the oscillating particle into a variety of non-equilibrium steady states with tuneable parameters, turning such nano-mechanical oscillators into ideal test-systems for studies of stochastic thermodynamics. Based on a Langevin equation written for the oscillating particle, we derive analytical expressions for the energy distribution in the stationary states and find that, under appropriate circumstances, our theoretical predictions agree very well with the energy distributions observed in the simulations.  In addition, we find that in our experiments parameter fluctuations dominate the noise contribution from Brownian motion, which leads to and additional broadening of the experimental distributions. 

In addition to the levitated nanoparticle considered here, our model applies to other nonlinear oscillators, including ultra high-Q nano-mechanical oscillators fabricated from silicon nitride \cite{Fong_2012} and carbon nanotubes and graphene resonators  \cite{Eichler_2011}. The latter naturally exhibit nonlinear damping that is formally identical to our feedback mechanism.
Thus, in addition to providing insights into thermodynamics on the nanoscale, the work presented here provides insight into the interaction of noise with inherent nonlinearities of nano-mechanical oscillators and the resulting amplitude and phase noise.
Most notably phase noise, despite being an active topic of research for many decades, is still a  pertinent topic today \cite{Bonnin_2013, Agrawal_2014, Moehlis_2014}, since it plays a prominent role for the application of such systems as sensors and in timing and frequency control.

The remainder of the article is organised as follows. In Sec. \ref{sec_theory} we lay out the theory for the energy distribution of a nano-mechanical oscillator subject to friction, nonlinear feedback cooling and parametric modulation. Computer simulations are then used, in Sec. \ref{sec_simulation}, to verify the theoretical predictions and probe the limits of the theory. In Sec. \ref{sec_experiment} we first describe our experimental setup and explain how we determine the relevant system parameters. We then present energy and phase distributions and discuss how they compare with theory and simulations. Some conclusions and an outlook are provided in Sec. \ref{sec_conclusion}.

\section{Theory}
\label{sec_theory}

\subsection{Equation of motion}

We consider a particle of mass $m$ oscillating in a trap with a Duffing potential 
\begin{equation}
V(q)=\frac{1}{2}kq^2 + \frac{1}{4}\xi k q^4,
\end{equation}
where $q$ specifies the position of the particle, $k$ is the trap stiffness,  and $\xi$ is the Duffing parameter,  which quantifies how strongly the trap deviates from a purely harmonic potential. Using the frequency $\Omega_0=\sqrt{k/m}$ of the harmonic case, the total energy of the oscillator is given by
\begin{equation}\label{eq:Energy}
E(q,p)=\frac{1}{2}m\Omega_0^2 q^2+\frac{1}{4}\xi m\Omega_0^2 q^4 + \frac{p^2}{2m},
\end{equation}
where $p=m\dot q$ is the momentum of the particle. The force due to the trap is hence given by 
\begin{equation}
F_{\rm trap}=-m\Omega_0^2 q-\xi m\Omega_0^2 q^3.
\end{equation}
Since the particle is immersed in a low density gas of temperature $T$, it experiences also a frictional force
\begin{equation}
F_{\rm friction}=-\Gamma p
\end{equation}
and the related fluctuating random force
\begin{equation}
F_{\rm random}=\sqrt{2m\Gamma k_{\rm B}T}\,w(t),
\end{equation}
where $\Gamma$ is the friction constant, $k_{\rm B}$ is the Boltzmann constant and $w(t)$ is white noise. A feedback of strength $\eta$, acting on the particle with force
\begin{equation}
F_{\rm feedback} = -\Omega_0 \eta q^2p,
\end{equation} 
used to control the effective temperature of the center of mass motion of the particle and cool it far below the gas temperature $T$ \cite{Gieseler_PRL2012}. In addition, the particle is driven parametrically by periodically modulating the trap stiffness with frequency $\Omega_m$ leading to the force 
\begin{equation}
F_{\rm drive}= \zeta m \Omega^2_0 \cos(\Omega_m t)q,
\end{equation}
where the modulation depth $\zeta$ determines the intensity of the parametric driving. Taken together, these forces yield the following stochastic equations of motion for the motion of the particle in the trap,
\begin{eqnarray}
{\rm d}q&=&\frac{p}{m}{\rm d}t, \label{equ:SDE1}\\
{\rm d}p &=&\left[-m\Omega_0^2q-\xi m\Omega_0^2 q^3-\Gamma p-\Omega_0 \eta q^2p+\zeta m \Omega^2_0 \cos(\Omega_m t)q\right]{\rm d}t \nonumber \\
& & +\sqrt{2m\Gamma k_{\rm B}T}\,{\rm d}W.
\label{equ:SDE2}
\end{eqnarray}
Here, $W(t)$ is the Wiener process with
\begin{eqnarray}
\langle W(t) \rangle & = & 0,\\
\langle W(t) W(t') \rangle & = & \min(t, t').
\end{eqnarray}
Note that $\langle W^2(t) \rangle = t$ for any time $t\ge 0$ and, thus, for an infinitesimal time interval ${\rm d}t$ one has $\langle ({\rm dW)^2}\rangle={\rm d}t$. The white noise $w(t)$ appearing in the random force can be viewed as the time derivative of the Wiener process, $w(t)= {\rm d}W(t)/{\rm d}t$.

In order to determine the energy distribution of the oscillator in the steady state, we now examine the time evolution of the energy generated by the stochastic equations of motion. To avoid multiplicative noise, i.e., a noise term with an amplitude depending on the current value of the energy, we consider the square root of the energy rather than the energy itself,
\begin{equation}
\epsilon(q, p)=\sqrt{E(q, p)}.
\end{equation}
Applying Ito's formula \cite{Gardiner_book} for the change of variables to $\epsilon(q, p)$ we find that the change ${\rm d}\epsilon$ during a short time interval is given by
\begin{equation}
{\rm d}\epsilon = \left[ \frac{pF(q,p,t)}{2m\epsilon}+\frac{\Gamma k_{\rm B}T}{2\epsilon}\left(1-\frac{p^2}{2m\epsilon^2}\right)\right] {\rm d}t+\sqrt{2m\Gamma k_{\rm B}T}\frac{p}{2m\epsilon}{\rm d}W,
\label{equ_depsilon}
\end{equation}
where
\begin{equation}
F(q,p,t)=-\Gamma p-\Omega_0 \eta q^2p+\zeta m \Omega^2_0 \cos(\Omega_m t)q
\end{equation}
is the sum of the non-conservative forces consisting of the frictional force $F_{\rm friction}$, the feedback force $F_{\rm feedback}$ and the driving force $F_{\rm drive}$. Note that the conservative forces, including the force due to the non-linear Duffing term in the energy, do not contribute to the energy change. 

The stochastic equation of motion for $\epsilon$, Equ. (\ref{equ_depsilon}), explicitly depends on the position and momentum of the particle. To eliminate this dependence and obtain a closed equation depending only on $\epsilon$,  we observe that the particle settles into a periodic motion with a frequency $\Omega$ that is not necessarily equal to the frequency $\Omega_0$ of the unperturbed oscillator. Integrating Equ. (\ref{equ_depsilon}) over one oscillation period $\tau=2\pi/\Omega$ we we obtain the change $\Delta \epsilon = \int_0^\tau {\rm d\epsilon}$ of $\epsilon$ during the time $\tau$,  
\begin{eqnarray}
\Delta  \epsilon &=& -\Gamma\int_0^\tau \frac{p^2}{2m\epsilon }{\rm d}t -\Omega_0\eta\int_0^\tau \frac{q^2p^2}{2m\epsilon }{\rm d}t  +
                                  \Gamma k_{\rm B}T\int_0^\tau \frac{1}{2\epsilon}\left(1-\frac{p^2}{2m\epsilon^2}\right){\rm d}t\nonumber\\
                          & &+ \zeta m \Omega^2_0 \int_0^\tau \frac{\cos(\Omega_m t)qp}{2m\epsilon}{\rm d}t  +  \sqrt{2m\Gamma k_{\rm B}T}\int_0^\tau \frac{p}{2m \epsilon }{\rm d}W.
\label{equ:delta_epsilon}
\end{eqnarray}
To compute the integrals, we assume that during this time, which at low friction is short compared to the time for energy relaxation, the particle performs an undisturbed harmonic oscillation evolving according to
\begin{equation}
q(t)=R \cos (\Omega t + \phi) \qquad \qquad p(t)=-m\Omega R \sin(\Omega t + \phi),
\end{equation}
where the amplitude $R$ of the oscillation is related to $\epsilon$ by $R=\sqrt{2/m}(\epsilon/\Omega)$. The phase $\phi$ accounts for a possible phase shift with respect to the driving force, which is proportional to $\cos(\Omega_m t)$. Note that the oscillation frequency $\Omega$ is not necessarily the same as the frequency $\Omega_0$ of the unperturbed harmonic oscillator. 

The central assumption, which allows to treat the motion of the system as that of an undisturbed oscillator during one oscillation period and eliminate the dependence on the rate of energy change on the phase space variables $q$ and $p$ by integration, is that the system evolves at nearly constant energy during one oscillation period. This condition is met if there is a separation of time scales between the time scale of the oscillation and the time scale for energy loss/gain. In other words, the relative change in energy $\Delta E / E$ occurring during one oscillation period should be much smaller than unity. The stochastic differential equation derived below for the time evolution of the energy, Equ. (\ref{equ:SDE_energy}), provides a way to estimate for which ranges of the parameters $\Gamma$, $\eta$ and $\zeta$ this condition holds.  Analyzing each term on the right hand side of Equ. (\ref{equ:SDE_energy}) individually, we find that the separation of time scale requires that $\Gamma / \Omega \ll 1$, $\zeta \ll 1$ and $\eta k_{\rm B}T_{\rm eff}/m\Omega^2 \ll 1$, where $T_{\rm eff}$ is the effective temperature of the oscillator.

Carrying out the integrals over $t$, the first three terms in Equ. (\ref{equ:delta_epsilon}) yield
\begin{equation}
\Delta  \epsilon' = -\frac{\Gamma \epsilon}{2} \tau
                           -\frac{\eta \epsilon^3 \Omega_0}{4m \Omega^2}\tau 
                           +\frac{\Gamma k_{\rm B}T}{4\epsilon}\tau 
\end{equation}
The change in $\epsilon$ resulting from the driving (fourth term in Equ. (\ref{equ:delta_epsilon})) is given by
\begin{equation}
\Delta  \epsilon'' = -\frac{\epsilon\zeta\Omega_0^2\sin(\pi\frac{\Omega_m}{\Omega} )\left[\cos(2\phi)\sin(\pi\frac{\Omega_m}{\Omega} )-(\frac{\Omega_m}{2\Omega}) \sin(2\phi)\cos(\pi\frac{\Omega_m}{\Omega}) \right]}{\Omega\pi\left(4-\frac{\Omega^2_m}{\Omega^2} \right)}\tau.
\end{equation}
This expression is independent of time only if after one oscillation period the relative phase of the oscillation with respect to the periodic driving force is the same as at the beginning of the period. For the parameters studied here and a modulation frequency of $\Omega_m \approx 2\Omega_0$, the oscillator locks to the modulation and oscillates with $\Omega=\Omega_m/2$. We limit our considerations to this case in the following. Carrying out the limit $\Omega_m \rightarrow 2\Omega$ in the above equation, one finds
\begin{equation}
\Delta  \epsilon'' = -\frac{\epsilon\zeta\Omega_0^2 \sin(2\phi)}{4\Omega}\tau.
\end{equation}
Finally, the last term in Equ. (\ref{equ:delta_epsilon}),
\begin{equation}
\Delta  \epsilon'''=\sqrt{2m\Gamma k_{\rm B}T_0}\int_0^\tau \frac{p}{2m \epsilon }{\rm d}W,
\end{equation}
is a stochastic integral due to the noise term in the equations of motion.  As a weighted sum of Gaussian random variables, $\Delta  \epsilon'''$ is also a Gaussian random variable with mean 
\begin{equation}
\langle \Delta  \epsilon''' \rangle = \sqrt{2m\Gamma k_{\rm B}T}\int_0^\tau \frac{p}{2m \epsilon }\langle {\rm d}W\rangle =0
\end{equation}
and variance 
\begin{eqnarray}
\langle (\Delta  \epsilon''')^2 \rangle &=& 2m\Gamma k_{\rm B}T\int_0^\tau\int_0^\tau \frac{p(t)p(t')}{4m^2 \epsilon^2 }\langle {\rm d}W{\rm d}W'\rangle\nonumber \\
& = & 2m\Gamma k_{\rm B}T\int_0^\tau \frac{p^2}{4m^2 \epsilon^2}{\rm d}t=\frac{\Gamma k_{\rm B}T}{2}\tau.
\end{eqnarray}
Thus, the random variable $\Delta  \epsilon'''$ can be written in terms of the Wiener process as
\begin{equation}
\Delta  \epsilon'''=\sqrt{\frac{\Gamma k_{\rm B}T}{2}} W(\tau).
\end{equation}
Putting things together, one obtains
\begin{equation}
\Delta  \epsilon = \left[-\frac{\Gamma \epsilon}{2} 
                           -\frac{\eta \epsilon^3 \Omega_0}{4m \Omega^2} 
                           +\frac{\Gamma k_{\rm B}T}{4\epsilon}
-\frac{\epsilon\zeta\Omega_0^2 \sin(2\phi)}{4\Omega}\right]\tau+ \sqrt{\frac{\Gamma k_{\rm B}T}{2}} W(\tau).
\end{equation}
Since the oscillation period $\tau$ is short compared to all the time scale on which the energy changes, one can finally write the following stochastic differential equation for the square root of the energy $\varepsilon$
\begin{equation}
{\rm d}  \epsilon = \left[-\frac{\Gamma \epsilon}{2} 
                           -\frac{\eta  \Omega_0 \epsilon^3}{4m \Omega^2} 
                           +\frac{\Gamma k_{\rm B}T}{4\epsilon}
-\frac{\epsilon\zeta\Omega_0^2 \sin(2\phi)}{4\Omega}\right]{\rm d}t+ \sqrt{\frac{\Gamma k_{\rm B}T}{2}} {\rm d}W
\label{equ_stochastic_epsilon}
\end{equation}
The corresponding Fokker-Planck equation \cite{Risken_book} governing the time evolution of the probability density function $P_\epsilon(\epsilon, t)$ is given by 
\begin{eqnarray}
\frac{\partial P_\epsilon(\epsilon, t)}{\partial t}&=&\frac{\partial }{\partial \epsilon}\left[\frac{\Gamma \epsilon}{2} 
                           +\frac{\eta  \Omega_0 \epsilon^3}{4m \Omega^2} 
                           -\frac{\Gamma k_{\rm B}T}{4\epsilon}
+\frac{\epsilon\zeta\Omega_0^2 \sin(2\phi)}{4\Omega}\right]P_\epsilon(\epsilon, t)\nonumber \\
&& +\frac{\Gamma k_{\rm B}T}{4}\frac{\partial^2 }{\partial \epsilon^2}P_\epsilon(\epsilon, t).
\end{eqnarray}
In writing these two equation we have implicitly assumed that the phase $\phi$ between the modulation and the particle oscillation is fixed (or at least that it changes only very slowly in time). As we will show below, this condition is met very well particularly at low friction. Equation (\ref{equ_stochastic_epsilon}) implies that the time evolution of $\varepsilon$ can be viewed as a Brownian motion in the high friction limit under the influence of an external force. Note that due to the integration over one oscillation period, this equation has $\epsilon$ as its only time dependent variable while the dependence on other variables has been removed. 
In the following section we will use this equation to determine the energy distribution as well as the phase space distribution of the steady state generated by the parametric modulation and the feedback mechanism.  

Changing variables from $\epsilon$ to $E=\epsilon^2$ and applying Ito's formula \cite{Gardiner_book} yields the corresponding stochastic differential equation for the energy,
\begin{equation}
{\rm d} E = \left[-\Gamma (E - k_{\rm B}T)
 -\frac{\eta  \Omega_0 E^2}{2m \Omega^2}
-\frac{E\zeta\Omega_0^2 \sin(2\phi)}{2\Omega}
\right]{\rm d}t 
+\sqrt{2E \Gamma k_{\rm B}T}{\rm d}W. 
\label{equ:SDE_energy}
\end{equation}
In contrast do stochastic equation of motion for $\epsilon$, here the noise is multiplicative, i.e., its amplitude is energy dependent. The corresponding Fokker-Planck equation for the probability density function $P_E(E, t)$ is given by
\begin{eqnarray}
\frac{\partial P_E(E, t)}{\partial t}&=&\frac{\partial }{\partial E}\left[\Gamma (E - k_{\rm B}T)
 +\frac{\eta  \Omega_0 E^2}{2m \Omega^2}
 +\frac{E\zeta\Omega_0^2 \sin(2\phi)}{2\Omega}\right]P_E(E, t)\nonumber \\
&& +{\Gamma k_{\rm B}T}\frac{\partial^2 }{\partial E^2}E P_E(E, t).
\end{eqnarray}

\subsection{Energy distribution}

The stochastic differential equation (\ref{equ_stochastic_epsilon}) has the form of the equation of motion describing the time evolution of a one-dimensional Brownian particle under the external force $f(x)$ with large friction $\nu$ at temperature $T$,
\begin{equation}
{\rm d} x = \frac{1}{\nu}f(x){\rm d}t+\sqrt{\frac{2k_{\rm B}T}{\nu}}{\rm dW},
\label{equ:high_friction}
\end{equation}
where $x$ is the position of the Brownian particle. The motion resulting from this equation of motion is known to sample the Boltzmann-Gibbs distribution
\begin{equation}
P_x(x) \propto \exp\left\{-\beta U(x)\right\},
\end{equation}
where $\beta=1/k_{\rm B} T$ is the reciprocal temperature and $U(x)$ is the potential corresponding to the external force, $f(x)=-{\rm d}U/{\rm d}x$.

By virtue of this isomorphism with over-damped Brownian motion, established by setting $\nu=4/\Gamma$ and identifying $\epsilon$ with $x$, the determination of the energy in the non-equilibrium steady state of the driven oscillator turns into an equilibrium problem.  One can then immediately infer that Equ. (\ref{equ_stochastic_epsilon}) samples the distribution 
\begin{equation}
P_\epsilon(\epsilon) \propto \exp\left\{-\beta U(\epsilon)\right\},
\end{equation}
where the potential 
\begin{equation}
U(\epsilon)=\epsilon^2 
+\frac{\eta  \Omega_0 \epsilon^4}{4m\Gamma \Omega^2} 
-k_{\rm B}T \ln \epsilon
+\frac{\epsilon^2\zeta\Omega_0^2 \sin(2\phi)}{2\Gamma\Omega}
\end{equation}
generates the force 
\begin{equation}
f(\epsilon)=-\frac{{\rm d} U(\epsilon)}{{\rm d}\epsilon}=-2\epsilon 
-\frac{\eta  \Omega_0 \epsilon^3}{m\Gamma \Omega^2} 
+\frac{k_{\rm B}T}{\epsilon}
-\frac{\epsilon\zeta\Omega_0^2 \sin(2\phi)}{\Gamma\Omega}
\end{equation}
acting on the variable $\epsilon$. As a result, the systems samples the $\epsilon$-distribution
\begin{equation}
P_\epsilon(\epsilon) \propto \epsilon \exp\left\{-\beta\left[\left(1+\frac{\zeta\Omega_0^2 \sin(2\phi)}{2\Gamma\Omega}\right)\epsilon^2
+\frac{\eta  \Omega_0}{4m\Gamma \Omega^2}\epsilon^4 \right]\right\}.
\end{equation}
Note that a small friction $\Gamma$ corresponds to large friction $\nu$ determining the time evolution of $\epsilon$ and, thus, the energy $E$ of the oscillator. By a change of variables from $\epsilon$ to $E$, we finally obtain the probability density function of the energy $E$,
\begin{equation}
P_E(E) = \frac{1}{Z} \exp\left\{-\beta\left[\left(1+\frac{\zeta\Omega_0^2 \sin(2\phi)}{2\Gamma\Omega}\right)E
+\frac{\eta  \Omega_0}{4m\Gamma \Omega^2}E^2 \right]\right\}.
\label{equ_energy_distribution}
\end{equation}
The normalisation factor $Z=\int P_E(E){\rm d}E$ is given by
\begin{equation}
Z = \sqrt{\frac{\pi m\Gamma \Omega^2}{\beta \eta  \Omega_0}}
 h \left(\sqrt{\frac{\beta m\Gamma \Omega^2}{\eta  \Omega_0}}\left(1+\frac{\zeta\Omega_0^2 \sin(2\phi)}{2\Gamma\Omega}\right) \right),
\end{equation}
where the function $h(x)$ is defined as
\begin{equation}
h(x)=\exp(x^2){\rm erfc}(x)
\end{equation}
and ${\rm erfc}(x)$ is the complementary error function. Thus, the energy distribution is that of an equilibrium system with effective energy
\begin{equation}
H=\left[1+\frac{\zeta\Omega_0^2 \sin(2\phi)}{2\Gamma\Omega}\right]E
+\frac{\eta  \Omega_0}{4m\Gamma \Omega^2}E^2
\end{equation}
and configurational partition function $Z$. While the term proportional to $E^2$ is caused by the feedback cooling, the term proportional to $E$ is affected only by the parametric modulation. 

According to Equ. (\ref{equ_energy_distribution}), the energy distribution is Gaussian with a cutoff at $E=0$. The maximum of the Gaussian is located at 
\begin{equation}\label{eqn:EnergyMean}
\bar E = -\frac{ 2m\Gamma \Omega^2}{ \eta  \Omega_0}\left[1+\frac{\zeta\Omega_0^2 \sin(2\phi)}{2\Gamma\Omega}\right]
\end{equation}
while its variance (neglecting the cutoff) is given by
\begin{equation}\label{eqn:EnergyVar}
\sigma^2_E=\frac{2m\Gamma \Omega^2k_{\rm B}T}{\eta  \Omega_0}.
\end{equation}
Hence, the width of the Gaussian does neither depend on the driving parameters nor on the phase $\phi$. 

\subsection{Phase space distribution}

Since for low friction the energy of the oscillator changes slowly, one can also obtain the full phase space density $P_{qp}(q,p)$ from the energy density $P_E(E)$. To determine the phase space density $P_{qp}(q,p)$, we consider the micro-canonical phase space distribution $P_{\rm mc}(q, p; \tilde E)$ of the oscillator evolving at a given constant total energy $\tilde E$,
\begin{equation}
P_{\rm mc}(q, p; \tilde E)=\frac{1}{g(\tilde E)}\delta [E(q, p)-\tilde E],
\label{equ_mc}
\end{equation}
where $\delta (x)$ is the Dirac delta function and we have denoted the fixed value of the energy with $\tilde E$ to distinguish it from the energy function $E(q, p)$, which depends on the position $q$ and the momentum $p$. The normalising factor $g(\tilde E)$ is the micro-canonical density of states, 
\begin{equation}
g(\tilde E) = \int {\rm d}q {\rm d}p\; \delta [E(q, p)-\tilde E].
\end{equation}
The phase space distribution of Equ. (\ref{equ_mc}) would be observed for an oscillator evolving freely in the absence of feedback and without coupling to a heat bath. Since for the parameter ranges studied here the energy is essentially constant over many oscillation periods, the total phase space density $P_{qp}(q,p)$ can be written by averaging the microcanonical distribution over the energy distribution,
\begin{equation}
P_{qp}(q, p)=\int {\rm d}\tilde E P_E(\tilde E) P_{\rm mc}(q, p; \tilde E) = \int {\rm d}\tilde E  \frac{P_E(\tilde E)}{g(\tilde E)}\delta [E(q, p)-\tilde E].
\end{equation}
This linear superposition of micro canonical distributions is valid as long as the energy changes slowly on the time scale of the oscillation period. For the low friction constants and the small feedback strength studied here this assumption is met even under non-equilibrium conditions.  Carrying out the integral yields
\begin{equation}
P_{qp}(q, p)=\frac{P_E[E(q,p)]}{g[E(q,p)]}.
\end{equation}
As further approximation, we now use the density of states $g(E)=2\pi/\Omega_0$ for the harmonic oscillator, thus neglecting the Duffing term of the potential in this part of the calculation, and obtain
\begin{equation}
P_{qp}(q, p)=\frac{\Omega_0}{2\pi}P_E[E(q,p)].
\end{equation}
Inserting the energy distribution from Equ. (\ref{equ_energy_distribution}) into this equation we finally find the phase distribution function
\begin{equation}
P_{qp}(q, p) = \frac{\Omega_0}{2\pi Z} \exp\left\{-\beta\left[\left(1+\frac{\zeta\Omega_0^2 \sin(2\phi)}{2\Gamma\Omega}\right)E(q,p)
+\frac{\eta  \Omega_0}{4m\Gamma \Omega^2}E(q,p)^2 \right]\right\}.
\label{equ_phase_space_distribution}
\end{equation}
Note, however, that while we have neglected the Duffing term in the expression for the density of states, it is included in the energy appearing in the argument of the exponential on the right hand side of the above equation. 

From the phase space density $P_{qp}(q,p)$ one can obtain the distribution $P_q(q)$ of the position by integration over the momenta, 
\begin{equation}
P_{q}(q)=\int_{-\infty}^{\infty}{\rm d}p \,P_{qp}(q,p).
\end{equation}
In the absence of parametric modulation ($\zeta=0$), one finds by carrying out the integral 
\begin{eqnarray}
P_q(q)&\propto& \sqrt{2+\eta\frac{\Omega_0}{\Gamma} \left(\frac{q^2}{2}+\xi \frac{q^4}{4}\right)} 
\exp \left(-\frac{\beta m \Omega_0}{8\eta} \left[2+\eta\frac{\Omega_0}{\Gamma} \left(\frac{q^2}{2}+\xi \frac{q^4}{4}\right)\right]^2\right)
\nonumber \\
& & \times K_{\frac{1}{4}}\left(\frac{\beta m \Omega_0}{8\eta} \left[2+\eta\frac{\Omega_0}{\Gamma} \left(\frac{q^2}{2}+\xi \frac{q^4}{4}\right)\right]^2\right),
\label{equ:distribution_position}
\end{eqnarray}
where $K_{1/4}$ is a generalised Bessel function of the second kind. For simplicity, we have considered the case $\Omega_0=\Omega$ here. A similar expression can also be derived for the momentum distribution. 

\subsection{Relative entropy change}

As shown recently, a fluctuation theorem holds for the relative entropy change $\Delta \mathcal{S}$ for a system relaxing towards equilibrium starting from the non-equilibrium steady state prepared by feedback cooling and parametric driving\cite{NAT_NANOTECH}. In this process, the feedback and the driving are turned off during the relaxation such that the system evolves freely and the dynamics is microscopically reversible. The relative entropy change $\Delta \mathcal{S}$ is defined as the logarithmic ratio of the probability $P[u(t)]$ to observe a certain trajectory $u(t)$ and the probability $P[u^*(t)]$ of the time reversed trajectory $u^*(t)$, 
\begin{equation}
\Delta \mathcal{S} = \ln \frac{P[u(t)]}{P[u^*(t)]}.
\end{equation}
Here, $u(t)$ denotes an entire trajectory of length $t$ including position and momentum of the oscillator and $u^*(t)$ denotes the trajectory that consist of the same states visited in reverse order with inverted momenta. Since during the relaxation detailed balance is obeyed, for the quantity $\Delta \mathcal{S}$ a detailed fluctuation can be proven, 
\begin{equation}
P_t(-\Delta \mathcal{S})/P_t(\Delta \mathcal{S}) = \exp(-\Delta \mathcal{S}),
\end{equation}
where $P_t(\Delta \mathcal{S})$ is the probability density to observe the value $\Delta \mathcal{S}$ at time $t$ as determined over many repetitions of the relaxation experiment. For the relaxation process considered in Ref. \cite{NAT_NANOTECH} the relative entropy change is given by 
\begin{equation}
\Delta \mathcal{S} = \beta Q_h + \Delta \phi,
\label{eq:deltaS}
\end{equation}
where $Q_h=-[E_t-E_0]$ is the energy absorbed by the bath during the relaxation, and $E_0$ and $E_t$ are the energy of the oscillator at time $0$ and $t$, respectively. The quantity $\phi(q, p)$ is defined as as the logarithm of the stationary phase space distribution
\begin{equation}
\phi(q, p) = -\ln P_{qp}(q,p)
\end{equation}
and $\Delta \phi$ is the difference of $\phi$ at the beginning and the end of the trajectory,
\begin{equation}
\Delta \phi =\phi_t-\phi_0.
\end{equation}
Hence, the relative entropy change  $\Delta \mathcal{S}$ depends on the state of the system at the beginning and the end of the trajectory.  

In general, the steady distribution $P_{qp}(q,p)$ necessary to compute $\Delta \phi$ is unknown. However, from the distribution derived for our model, Equ. (\ref{equ_phase_space_distribution}), we find that for the relaxation from a non-equilibrium steady state generated by nonlinear feedback and parametric modulation, the relative entropy change is given by
\begin{equation}
  \Delta \mathcal{S}=
 -\beta \frac{\zeta\Omega_0^2 \sin(2\phi)}{2\Gamma\Omega}\left[E_t-E_0\right]
 -\beta \frac{\eta  \Omega_0}{4m\Gamma \Omega^2}\left[E^2_t-E^2_0\right],
\end{equation}
Thus, our stochastic model allows us to express the relative entropy change during a relaxation trajectory in terms of the energy at the beginning and the end of that trajectory. Note that since no work is performed on the system, the heat $Q_h$ exchanged along a trajectory  equals the energy lost by the system. Thus, in the absence of nonlinear feedback cooling, the relative entropy change is proportional to the heat and resembles relaxation form a thermal bath with effective temperature $T_{\rm eff} = T/(1-\frac{\zeta\Omega_0^2 \sin(2\phi)}{2\Gamma\Omega})$. By choosing parameters, one can therefore switch from a purely thermal situation with the phase space distribution of a harmonic oscillator (but with changed temperature) to a truly non-equilibrium steady-state with non-linear effects controlled by the feedback parameter $\eta$. 

\subsection{Quadratures}

Parametrically driven nano-mechanical oscillators have been shown to support classical squeezed states in which the amplitude of the vibration in one phase is reduced with respect to the thermal equilibrium amplitude. To probe our oscillator for squeezed states we analyse its motion in terms of the so-called quadratures. For the oscillator driven by the parametric modulation $F_{\rm drive}= \zeta m \Omega^2_0 \cos(\Omega_m t)q$, we write the time evolution of the oscillator position as
\begin{equation}
q(t)=R(t) \cos[\Omega t + \phi(t)],
\label{equ_qoft}
\end{equation}
where $\Omega$ is the frequency of the particle oscillating at half the frequency of the driving, $\Omega=\Omega_m /2$. Here, $R(t)$ and $\phi(t)$ are the amplitude and the phase of the particle, respectively, and the phase is measured with respect to the driving signal. Using the addition theorem for the sine-function, Equ. (\ref{equ_qoft}) can be written as the sum of two contributions, one in-phase with the driving signal and one out-of-phase,
\begin{eqnarray}
q(t)&=&R(t) \cos \phi(t) \cos (\Omega t) - R(t) \sin \phi (t) \sin(\Omega t)\nonumber \\
&=&X(t) \cos (\Omega t) - Y(t)\sin(\Omega t),
\label{equ_qoft_XY}
\end{eqnarray}
where the second line defines the {\em in-phase} component $X(t) =R(t) \cos \phi(t)$ and the {\em quadrature} $Y(t)=R(t) \sin \phi (t)$. Together, $X$ and $Y$ are referred to as the quadratures.
The quadratures can be computed from the time evolution of the position $q(t)$ and the momentum $p(t)$.
The momentum of the particle is given by:
\begin{eqnarray}
p(t)/m\Omega&=&-R(t) \sin [\Omega t + \phi(t)] \nonumber \\
&=&-X(t) \sin (\Omega t) - Y(t)\cos(\Omega t),
\label{equ_poft_XY}
\end{eqnarray}
where we neglected the time derivatives of the amplitude and phase, since for an oscillator at low friction both the amplitude and the phase vary slowly in time.
Combining this equation with Equ. (\ref{equ_qoft_XY}) yields
\begin{eqnarray}
X(t)=& \;\;\;\; q(t) \cos (\Omega t) - \frac{p(t)}{m\Omega}\sin (\Omega t),\nonumber \\
Y(t)=&-q(t) \sin (\Omega t) - \frac{p(t)}{m\Omega}\cos (\Omega t).
\label{equ_XY}
\end{eqnarray}
corresponding to transformation to a coordinate system that rotates clockwise with frequency $\Omega$  with respect to the $(q, p/m\Omega)$-plane \cite{Holmes_Rand_1976,Hale1969}. In this coordinate system, a sinusoidal oscillation of frequency $\Omega$ is represented by a static point.

Note that the amplitude and phase can be expressed in terms of the quadratures,
\begin{eqnarray}
R & = & \sqrt{X^2+Y^2}, \nonumber \\
\phi &= &\arctan(Y/X),
\end{eqnarray}
and that
\begin{equation}\label{eq:EofR}
\frac{m\Omega^2R^2}{2}=\frac{m\Omega^2}{2}(X^2+Y^2)
\end{equation}
is the energy of a harmonic oscillator with frequency $\Omega$. 

\section{Simulations}
\label{sec_simulation}

In this section we verify the analytical expressions for the distributions of energy and positions by comparing them with simulation results. The simulations were performed for parameter values close to those of the experiments, which we will present and discuss subsequently. 

\subsection{Simulation methods}

In our simulations, we integrated the Langevin equation of motion with the OVRVO algorithm of Sivak, Chodera and Crooks \cite{Sivak}, which can be viewed as a stochastic generalisation of the velocity Verlet algorithm for deterministic dynamics \cite{Book_Allen}. This discrete time integration scheme uses a time step rescaling in the deterministic update step for positions and momenta to satisfy a number of desiderata proposed in the literature for stochastic integrators \cite{Pastor}. In all simulations we used a time step of $\Delta t=0.01$ in reduced units. This time step is about 1/628 of the oscillation period. Test runs carried out with smaller time steps ($\Delta t=0.001$) yielded identical results up to statistical errors. In most cases, the total simulation time was $t=10^7$ corresponding to about $3\times 10^6$ modulation cycles. For some parameters we carried out longer simulations of up to $3\times 10^{10}$ steps corresponding to a total simulation time of $t=3\times 10^8$. All simulations were carried out for $k_{\rm B}T= 1$, $m=1$, and $k=1$.


To facilitate comparison of the results of theory/simulation and experiments, in the following we use the thermal energy $\mathcal{E}=k_{\rm B}T$, the inverse frequency $\mathcal{T}=1/\Omega_0$ and the particle mass $\mathcal{M}=m$ as our basic units of energy, time and mass, respectively. Accordingly, distances are measured in units of $\mathcal{L}=(1/\Omega_0)\sqrt{k_{\rm B}T/m}$ and velocities in units of $\mathcal{V}=\sqrt{k_{\rm B}T/m}$. Hence, the unit of length is given by the variance of the position of the harmonic oscillator, $\langle q^2 \rangle = k_{\rm B}T/m\Omega_0^2=\mathcal{L}^2$ and the unit of energy is the average energy of the harmonic oscillator $\langle E \rangle = k_{\rm B}T= \mathcal{E}$. The friction constant is given in units of $\Omega_0$ such that it equals the inverse of the quality factor, $Q=\Omega_0/\Gamma=1/\Gamma \mathcal{T}$. The feedback strength $\eta$ and the Duffing coefficient $\xi$ have the dimension of $1/$area and are measured in units of $1/ \mathcal{L}^2$. The modulation depth $\zeta$ is dimensionless. In the following, we use reduced units in which $\mathcal{E}=\mathcal{T}=\mathcal{M}=1$. 

\subsection{Oscillator with feedback cooling but without parametric modulation}

\begin{figure}[htb]
 \centerline{\includegraphics[height=4.5cm]{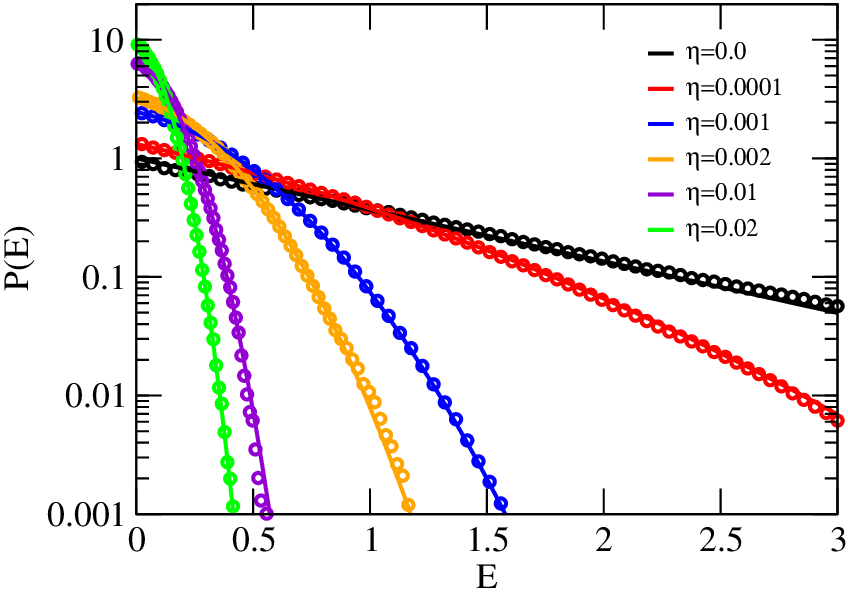}
                   \hspace{0.2cm}
                   \includegraphics[height=4.5cm]{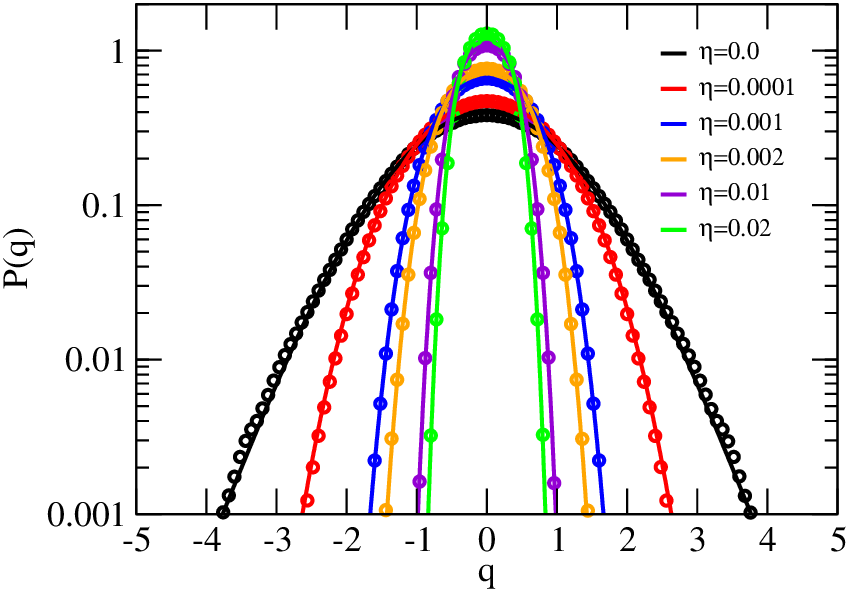}}
 \caption{{\em Left:} Energy distributions for different feedback strengths $\eta$ without parametric modulation ($\zeta=0$) for $\Gamma=0.0001$, and $\xi=-0.022$. The symbols are simulation results and the lines predictions according to Equ. (\ref{eq:EnergyDist_FBon}). {\em Right:} Position distributions for the same parameters. The symbols are simulation results and the lines are theoretical predictions according to Equ. (\ref{equ:distribution_position}).}
\label{fig:PE_nomodulation}
\end{figure}

We first consider the oscillator without parametric modulation ($\zeta=0.0$) but subjected to feedback cooling. Without driving, the phase $\phi$ is not a relevant parameter and the expression for the energy distribution simplifies considerably,
\begin{equation}\label{eq:EnergyDist_FBon}
P_E(E) \propto \exp\left\{-\beta\left(E+\frac{\eta}{4m\Gamma \Omega_0}E^2 \right)\right\},
\end{equation}
where we have assumed that the particle oscillates with $\Omega=\Omega_0$. The first term in the exponential is the same as that of the uncooled oscillator, but the second term proportional to $E^2$ is due to the feedback loop and strongly penalises high energy states thereby cooling the system. The cooling effect is stronger for weak friction $\Gamma$ and small frequencies $\Omega_0$. Several energy distributions obtained from simulations together with the corresponding predictions of Equ. (\ref{eq:EnergyDist_FBon}) are shown in the left panel of Fig. \ref{fig:PE_nomodulation}. The simulations were carried out for a friction of $\Gamma=0.0001$ and a Duffing parameter of $\xi=-0.022$.
Without feedback, $\eta=0$, the energy distribution is exponential, but for $\eta > 0$ the $E^2$ term caused by the feedback suppresses high energies leading to a parabolic shape of the distribution in the logarithmic representation. In all cases, the theoretical predictions agree very well with the simulation results. Positions distributions for the same set of parameters are shown in the right panel of Fig. \ref{fig:PE_nomodulation}. While without feedback the position distribution is Gaussian, the feedback quenches large deviations leading to a narrowing of the distributions. Also in the case of the position distributions the agreement between theory and simulation is excellent. 

\subsection{Oscillator with parametric modulation but without feedback cooling}

\begin{figure}[htb]
 \centerline{\includegraphics[height=4.5cm]{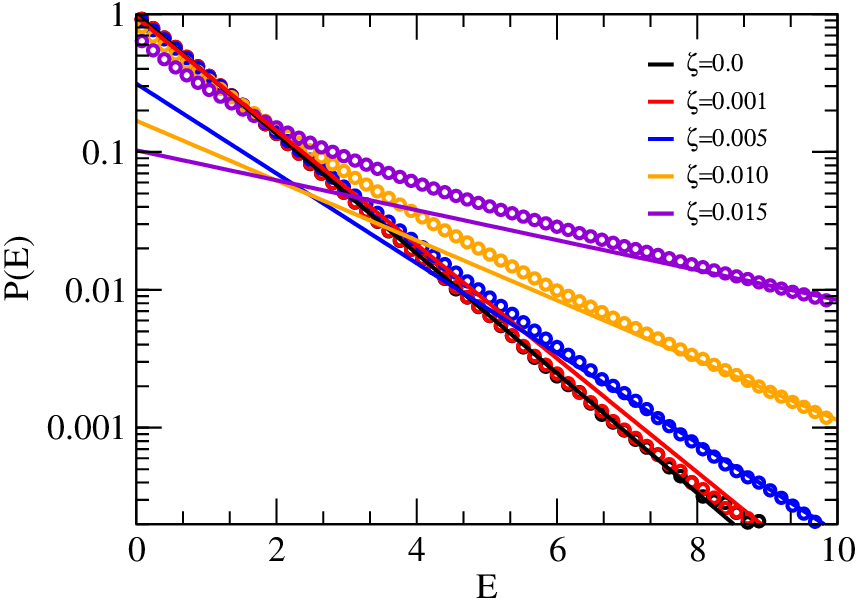}
                   \hspace{0.2cm}
                   \includegraphics[height=4.5cm]{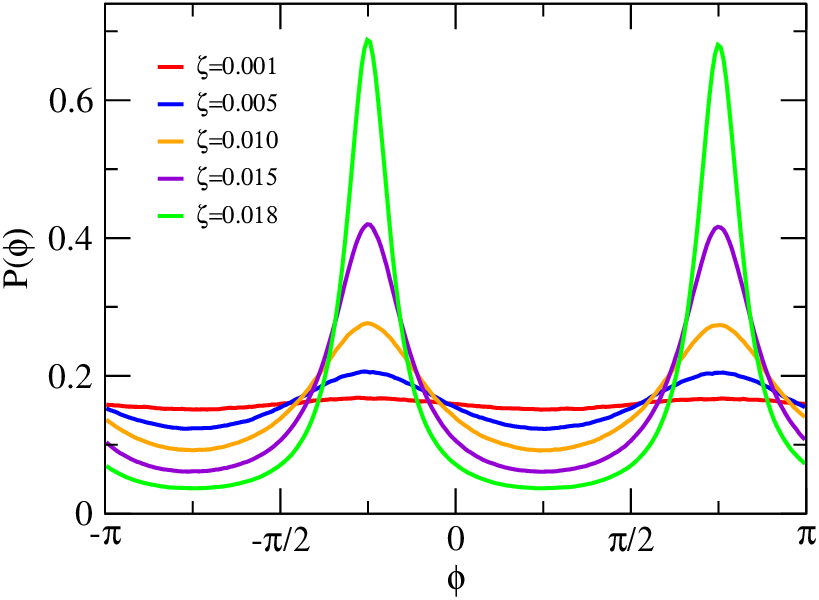}}
 \caption{{\em Left:} Energy distributions for different modulation depths $\zeta$ without feedback cooling ($\eta=0$) for $\Gamma=0.01$, $\xi=0$, $k_{\rm B}T= 1$, $m=1$, $k=1$, and $\Omega_m=2\, \Omega_0$. The symbols are simulation results and the lines predictions of the theory. The theoretical predictions have been scaled by a factor such that they agree with the numerical results at high energies.  {\em Right:} Phase distributions for different modulation depths $\zeta$ obtained from the same simulations. }
\label{fig:PE_nofeedback}
\end{figure}

We next turn to the oscillator with parametric driving but without feedback cooling. In this case, the energy deposited in the system by the modulation is removed only by the coupling to the gas as quantified by the friction constant $\Gamma$. If the particle oscillation is locked to the driving with a fixed phase $\phi$, the resulting energy distribution following from Equ. (\ref{equ_energy_distribution}) is expected to be exponential,
\begin{equation}
P_E(E) \propto
\exp\left\{-\beta\left(1+\frac{\zeta\Omega_0 \sin(2\phi)}{2\Gamma}\right)E \right\},
\end{equation}
where we have assumed that the modulation frequency is $\Omega_m=2\Omega_0$. For a vanishing Duffing parameter $\xi=0.0$, i.e., for a perfectly harmonic trap, the phase is expected to be $\phi=-\pi/4$ in the absence of thermal fluctuations \cite{LifshitzCross2008}. If this is the case, the decay constant of the exponential is $\beta (1-\zeta \Omega_0/2\Gamma)$. Hence, the decay constant is positive only for $ \zeta < 2\Gamma/\Omega_0$. If the modulation depth $\zeta$ exceeds this limit, the friction cannot remove the energy pumped into the oscillator by the modulation such that the oscillator energy keeps growing preventing the system from settling in a steady state. We indeed find in our simulations that for $ \zeta > 2\Gamma/\Omega_0$ the energy continuously increases. For weak driving, on the other hand, the energy distribution is expected to be exponential with the decay constant predicted by Equ. (\ref{equ_energy_distribution}). Several energy distributions for this case are shown in Fig. \ref{fig:PE_nofeedback}. Note that we performed these calculations for a relatively large friction constant of $\Gamma=0.01$, because for lower friction it takes exceedingly long to sample all relevant energies. For weak driving, $\xi=0.001$ (red symbols), the energy distribution is exponential as predicted by the theory. The negative slope of this distribution in the logarithmic representation is, however, slightly too large. The reason for this discrepancy is that the oscillation does not lock to the parametric driving as can bee seen in the distribution of the phase $\phi$ shown in the right panel of Fig. \ref{fig:PE_nofeedback}. The theory developed above, on the other hand, assumes a fixed phase of $\phi=-\pi/4$ (for $\xi=0$). For $\xi=0.001$, the phase distribution is essentially flat implying that there is no preferred phase. As a consequence, essentially no heating occurs and the energy distribution is indistinguishable from the equilibrium distribution (black symbols). As the strength of the parametric driving is increased, a pronounced phase relation between driving and oscillation develops and two distinct peaks appear in the phase distribution at equivalent positions, one at $\phi=-\pi/4$ and one at $\phi=-\pi/4+\pi$. Since the phase relation is more pronounced at high energies, in this regime the energy distributions shown in the left panel of Fig. \ref{fig:PE_nofeedback} converge to the form predicted by theory. In the figure, the theoretical distributions are indicated by lines with logarithmic slope of  $-\beta (1-\zeta \Omega_0/2\Gamma)$. For low energies, the phase relation is lost and the energy distributions have the logarithmic slope of the equilibrium distribution. Thus, the energy injected into the system by the parametric driving results in a longer tail in the energy distribution where it has the right phase relationship with the oscillation. In contrast at low energies, the form of the distribution is essentially unchanged with respect to the equilibrium distribution. 

\subsection{Oscillator with feedback cooling and parametric driving}

\begin{figure}[htb]
 \centerline{\includegraphics[width=12.0cm]{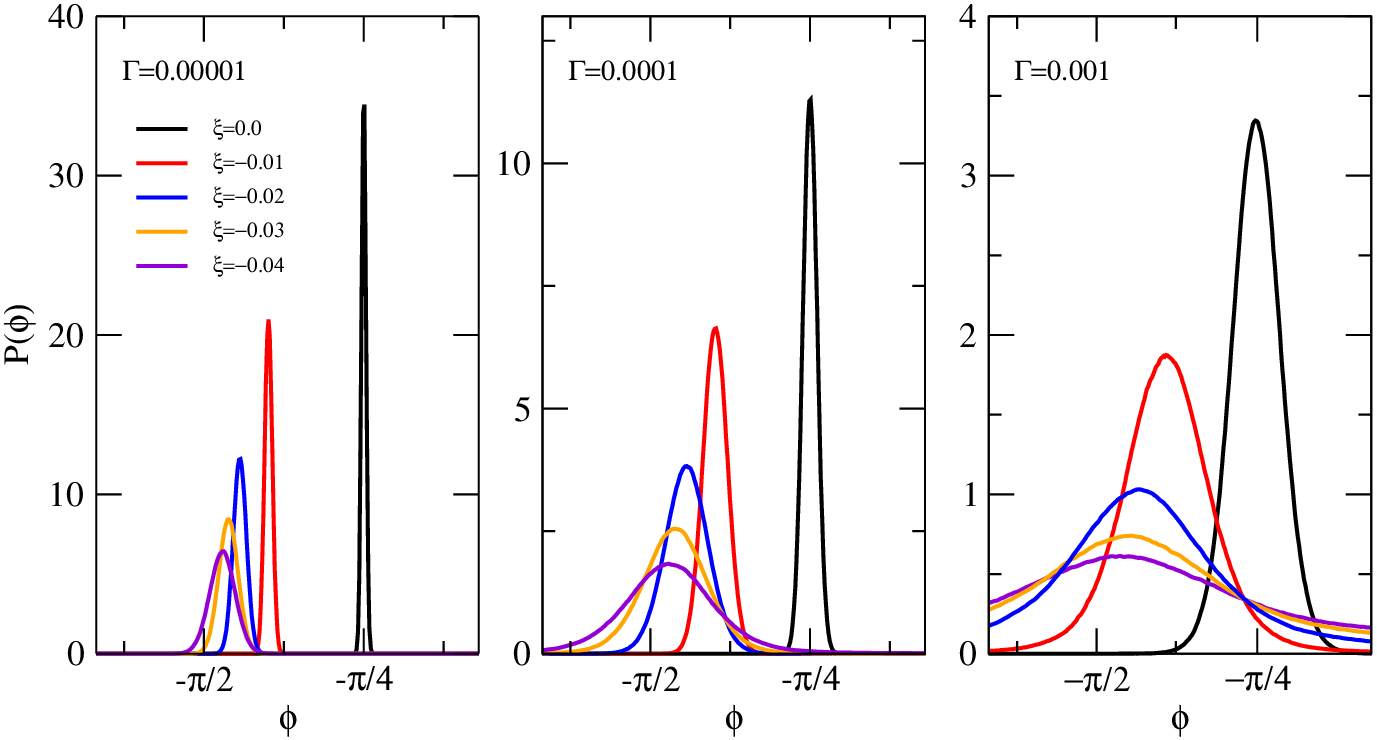}}
 \caption{Distributions of the phase $\phi$ for friction constants $\Gamma=0.00001$, 0.0001 and 0.01 and for different Duffing parameter $\xi$. The simulations were carried out for  $\eta= 0.022$, $\zeta=0.03$ and $\Omega_m=2\, \Omega_0$. }
\label{fig:Pphi}
\end{figure}

Next, we consider the oscillator with parametric driving {\em and } feedback cooling. To understand the energy distributions for this case, we first take a closer look at the statistics of the phase $\phi$. In the derivation of the analytical energy distribution, Equ. (\ref{equ_energy_distribution}), we have assumed a fixed phase $\phi$ between the modulation and the particle oscillation. In practice, however, the phase $\phi$ follows a statistical distribution with a position and width that depend on the parameters, particularly on the Duffing parameter $\xi$ and the friction constant $\Gamma$. Several distributions of the phase obtained from our simulations for $\Gamma$ and  $\xi$ are shown in Fig. \ref{fig:Pphi}. These simulations were carried out for a modulation depth of $\zeta=0.03$ and and a feedback strength of $\eta=0.022$, because these values can be realised in experiments. For all parameters considered here, the phase distributions are strongly peaked at a particular phase. The peaks are narrow for small friction and small Duffing parameters and broaden for increasing friction and non-linearity. Note that the Duffing parameters considered here are negative because the non-linearity is due to the shape of the focal intensity distribution, which is approximately Gaussian \cite{Gieseler_NATPHYS}. Without non-linearity, $\xi=0.0$, the peak is located at $\phi=-\pi/4$ for all values of the friction constant. As one turns on the non-linearity by making the Duffing parameter more negative, the peaks become broader and shift towards more negative values. 

\begin{figure}[tb]
 \centerline{\includegraphics[height=4.5cm]{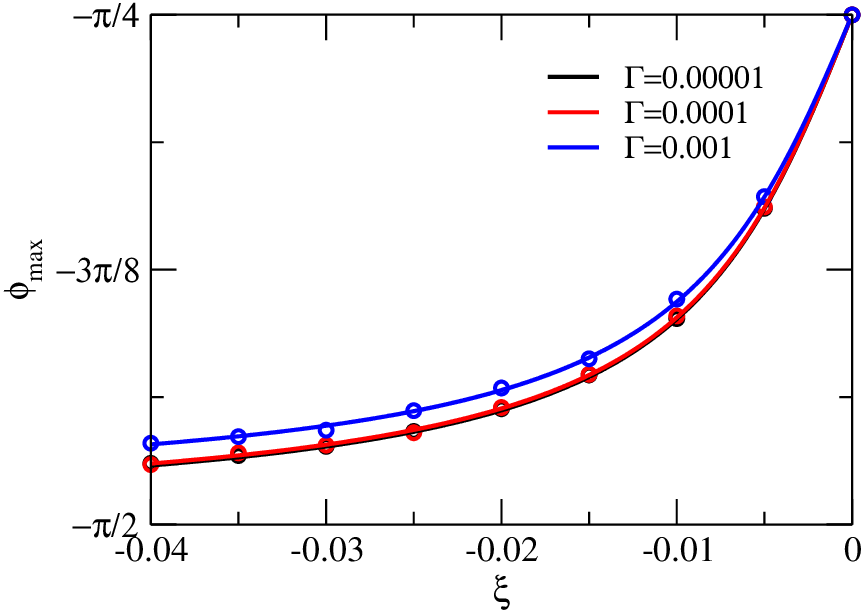}
                   \hspace{0.2cm}
                   \includegraphics[height=4.5cm]{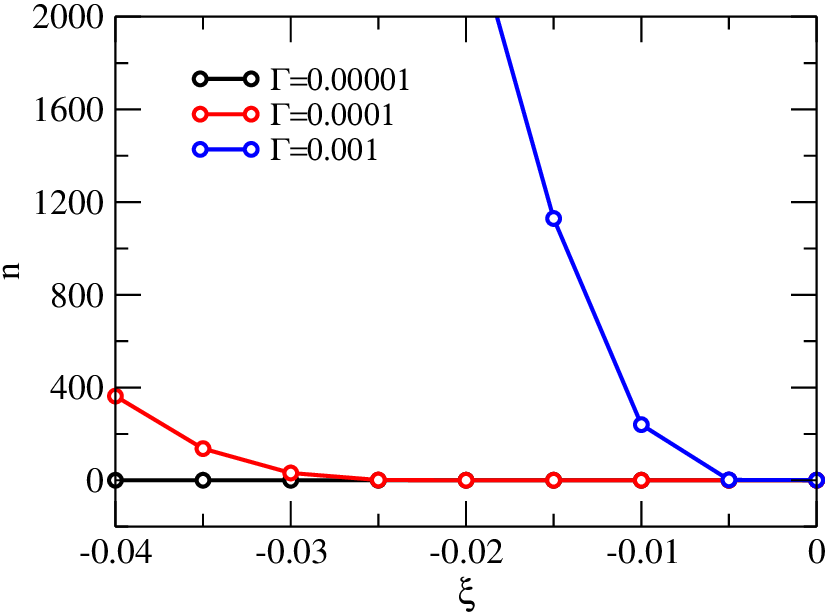}}
 \caption{{\em Left:} Most probable phase $\phi_{\rm max}$ as a function of the Duffing parameter $\xi$ for the friction constants $\Gamma=0.00001$ (black), $\Gamma=0.0001$ (red) and $\Gamma=0.001$ (blue). The simulations were carried out for $\eta= 0.022$, $\zeta=0.03$ and $\Omega_m=2\, \Omega_0$. The symbols are simulation results and the lines are results of secular perturbation theory. {\em Right:} Number of full turns the oscillation fell behind the driving during the total simulation time of $t=10^7$ as a function of the Duffing parameter  $\xi$ for the friction constants $\Gamma=0.00001$ (black), $\Gamma=0.0001$ (red) and $\Gamma=0.001$ (blue).}
\label{fig:phimax}
\end{figure}

A closer analysis of how the phase depends on the Duffing parameter is shown in Fig. \ref{fig:phimax}. The left panel of the figure shows the positions of the maximum of the phase distribution. i.e., the most likely phase $\phi_{\rm max}$, as a function of the Duffing parameter $\xi$ for different friction constants $\Gamma$. As can be inferred from the figure, the most likely phase $\phi_{\rm max}$ determined from the simulations (symbols) follows exactly the form predicted by secular perturbation theory \cite{LifshitzCross2008} (solid lines). While this theory neglects thermal fluctuations and cannot predict the entire phase distribution, it yields an accurate location of the maximum. 

Due to the thermal fluctuations, which lead to a broadening of the phase distribution, the oscillator might entirely loose the lock with the driving modulation and regain it only after falling behind by one entire turn of $2\pi$. For the lowest friction studied here this never happens during a simulation of total time $t=10^7$, but for higher frictions, and in particular for large Duffing parameters, the oscillation may fall behind the parametric modulation several times. The number of times this occurs in the course of the simulations is shown in the right panel of Fig. \ref{fig:phimax} for different friction constants as a function of $\xi$. 

\begin{figure}[tb]
 \centerline{ \includegraphics[height=4.5cm]{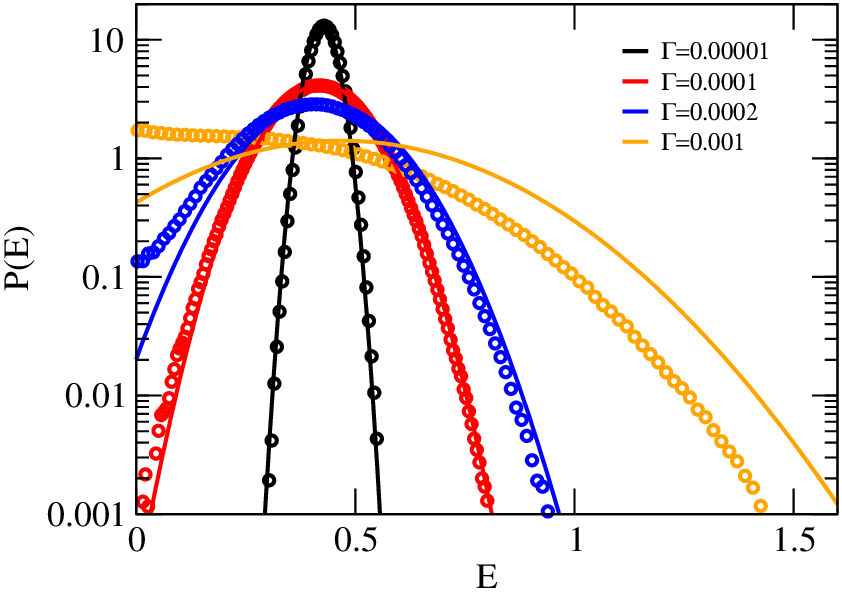}
                  \hspace{0.2cm}
                  \includegraphics[height=4.5cm]{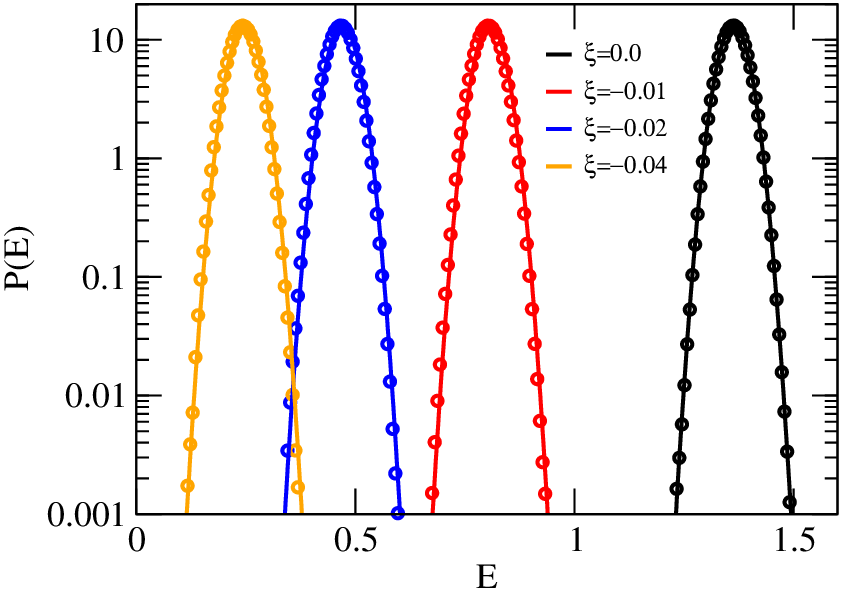}
 }
 \caption{{\em Left:} Energy distributions for different friction constants $\Gamma$, for $\xi=-0.022$, $\eta= 0.022$, $\zeta=0.03$ and $\Omega_m=2\, \Omega_0$. The symbols are simulation results and the lines predictions of the theory. {\em Right:} Energy distributions for different Duffing parameters $\xi$ for t $\Gamma=0.00001$, $\eta= 0.022$, $\zeta=0.03$ and $\Omega_m=2\, \Omega_0$. The symbols are simulation results and the lines predictions of the theory.}
\label{fig:PE_friction_duffing}
\end{figure}

We now compare the energy distribution determined in our simulations for the oscillator with parametric driving and feedback cooling with the theoretical prediction of Equ. (\ref{equ_energy_distribution}). To do that, we identify the phase $\phi$ occurring in the theoretical expression with the most likely phase $\phi_{\rm max}$ determined in the simulations. Energy distributions obtained for friction constants ranging from $\Gamma=10^{-5}$ to $\Gamma=10^{-3}$ are shown in the left panel of Fig. \ref{fig:PE_friction_duffing}. In all cases, the system was driven at $\Omega_m=2\Omega_0$ and the Duffing parameter, the feedback strength and the modulation depth were $\xi=-0.022$, $\eta= 0.022$, $\zeta=0.03$, respectively. While for high friction the theoretical predictions deviate considerably from the energy distributions determined in the simulations, most likely due to the lack of a stable phase relation, very good agreement is obtained for low friction, where phase distributions are strongly peaked.  This excellent correspondence is confirmed by the energy distributions shown along with theoretical predictions in the right panel of Fig. \ref{fig:PE_friction_duffing} for different Duffing parameters at low friction. Thus, the position and the width of the energy distribution in the non-equilibrium steady state generated by driving and cooling at the same time can indeed be controlled independently by an appropriate choice of parameters. 

\subsection{Quadratures}

\begin{figure}[htb]
\center
\hspace{1cm}
 \includegraphics[width=4.cm]{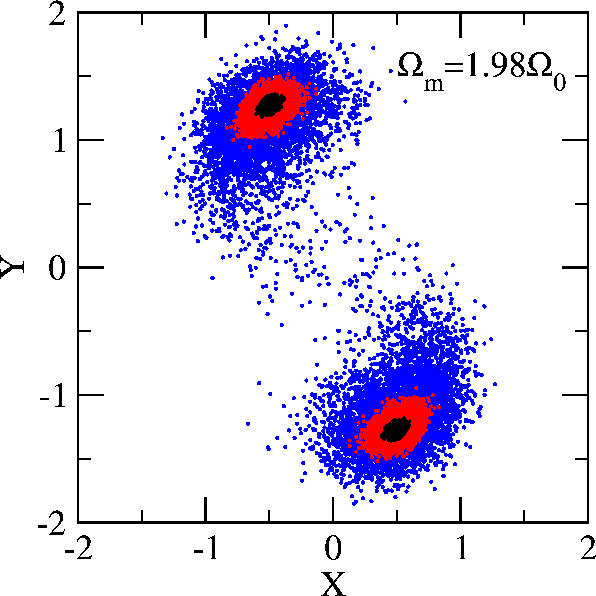}
 \hspace{0.0cm}
 \includegraphics[width=4.cm]{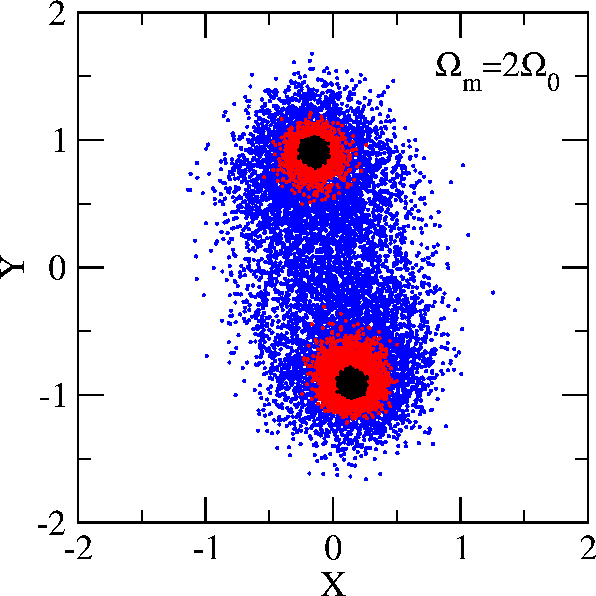}
 \hspace{0.0cm}
 \includegraphics[width=4.cm]{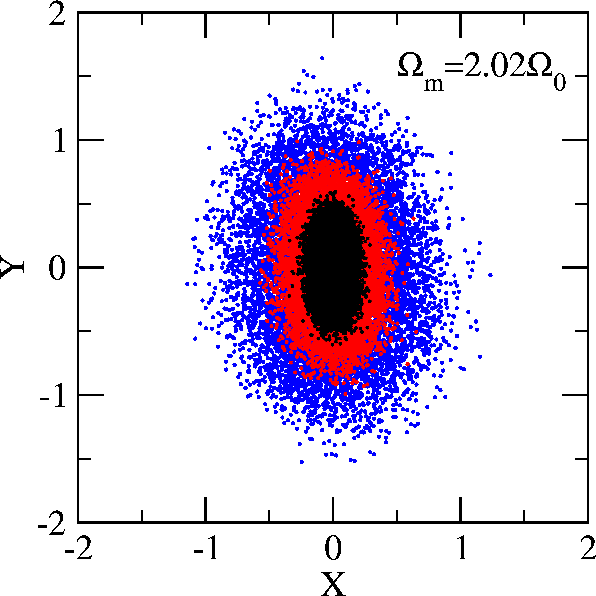}
 
 \caption{Scatter plot of the quadratures $X$ and $Y$ for the friction constants $\Gamma=0.00001$ (black), 0.0001 (red) and 0.01 (blue) for $\Omega_m=1.98\, \Omega_0$ (left), $\Omega_m=2\, \Omega_0$ (center) and $\Omega_m=2.02\, \Omega_0$ (right). The simulations were carried out for $\eta= 0.022$, $\xi=-0.022$, and $\zeta=0.03$.}
\label{fig:XY}
\end{figure}

Finally, we take a look at the distribution of the quadratures $X$ and $Y$ for different driving frequencies. Scatter plots of the quadratures obtained at different driving frequencies and for different values of the friction constant are shown in Fig. \ref{fig:XY}. From left to right, the driving frequency $\Omega_m$ is slightly below $2\Omega_0$, equal to $2\Omega_0$ and slightly above $2\Omega_0$. As in previous simulations, the parameters were $\xi=-0.022$, $\eta= 0.022$, and $\zeta=0.03$. At $\Omega_m=2\Omega$ and low friction the quadratures of the driven system are Gaussian with equal width along the two quadrature axes. Thus, they resemble a thermal state, albeit, displaced from the origin.
In contrast, for driving frequencies off $2\Omega_0$, the distributions are deformed, indicating the occurrence of classical squeezing.

\section{Experiments}
\label{sec_experiment}

In this section, we discuss how to retrieve the energy and phase of a trapped nanoparticle from discrete measurements of the particle positions.
From the retrieved energies and phases we reconstruct the energy and phase distributions and compare them to the theory and simulation results presented in the previous sections.
This allows us to extract the experimental parameters, which are detailed in Table \ref{tab:ExperimentalParameter}.
While the maxima of the distributions are in good agreement with our theory and simulations, the width of the experimental distributions is significantly broader due to parameter fluctuations not taken into account in the theoretical considerations.

\subsection{Experimental configuration}

In our experiments we use a silica nanoparticle trapped at the focus of a single beam optical tweezers.
The optical tweezer is formed by a $1064\,\rm nm$ laser beam ($\sim 35 \rm mW$) focused by a $\rm NA = 0.9$ objective, which is mounted inside a vacuum chamber.
The particle motion is recorded with an additional colinear laser ($780\rm nm$) and three balanced photodetectors.
A home-built electronic circuit is used to generate the feedback signal ($\eta$), while a frequency generator serves as the parametric modulation signal ($\zeta$).
The approximately Gaussian shape of the optical potential is responsible for the trap anharmonicity ($\xi$) \cite{Gieseler_NATPHYS}.
The detectors and the size of the nanoparticle are calibrated from measurements of the power spectral density of the particle motion at $5.1\rm mBar$.
At this pressure the $Q$-factor is high enough to resolve the three spatial modes, while broadening effects due to nonlinear mode coupling are negligible \cite{Gieseler_NATPHYS}.
For further details of the experimental configuration and calibration procedure see Refs. \cite{Gieseler_Thesis, Gieseler_PRL2014}.
Subsequent measurements are carried out at $1.2\times 10^{-5}\rm mBar$.

While our theoretical model is one-dimensional, the particle in the experiment moves in three dimensions along three main axes. The three axes  are determined by the symmetry of the laser focus. However, there is no direct coupling between the three spatial modes. In addition, feedback cooling reduces the amplitude such that also the nonlinear coupling becomes very weak. Therefore, our one-dimensional model is a very good approximation for the particle motion along one of the three main axes.

\begin{table}[tb]
\begin{center}
\begin{tabular}{|c|c|c|c|}

\hline
Parameter				        & Value  (phys. units)                        & Value (dimension less)    & error (\%)\\
\hline
a                             & $82 \pm 4$ nm                                 & $2.3 \times \mathcal{L}$                         & 4\\
m                             & $5.2\pm 0.7 \times 10^{-18}$ kg               & $1\times \mathcal{M}$                         & 13 \\
$\eta$                        &  $3.9 \pm 1.3 \, \mu m^{-2}$                  & $4.9\times 10^{-3}\times \mathcal{L}^{-2}$       & 34\\
$\xi$                         &  $-5.4\pm 1.1\, \mu m^{-2}$                  & $-6.9\times 10^{-3}\times \mathcal{L}^{-2}$      & 20\\
$\Gamma$                    &  $2\pi \times 8.1\pm0.2 \times 10^{-3}$ Hz    & $6.25\times 10^{-8}\times \mathcal{T}^{-1}$      & 3\\
$Q$                           &  $1.54\pm0.03 \times 10^7$                    & $1.54\times 10^7$         & 3\\
$\Omega_0     $               &  $2\pi \times 125.12 \pm 0.05$  kHz           & $1\times \mathcal{T}^{-1}$                         & 0.04\\
$\zeta$                       &  $16.1 \pm 1.3 \times 10^{-3}$                    & $16.1  \times 10^{-3}$      & 36\\
\hline
\end{tabular}
\caption{Overview of experimental parameters.
The second column lists the parameter in SI units with their respective experimental uncertainties, while the third column shows the experimental parameters in dimensionless units. For the scaling to dimensionless units see section 3.1. The last column lists the relative uncertainty of the experimental parameters.
\label{tab:ExperimentalParameter}
}
\end{center}
\end{table}

\subsection{Amplitude and phase estimation}

The particle oscillation frequencies along the three main axes are well separated and don´t overlap. Therefore, we can apply the maximum likelihood estimation for a single tone signal, that is a signal containing only one frequency component. The maximum likelihood estimation of the oscillation amplitude and phase of a single tone signal $q(t)$ is given by \cite{Rife1974} 
\begin{eqnarray}\label{eq:aML}
R_{\rm ML} &= |A_q(\Omega)|\\ \label{eq:thetaML}
\phi_{\rm ML} &= {\rm arg}\left[ \exp(-i\Omega_0 t_0  )  A_q(\Omega ) \right]
\end{eqnarray}
where $\Omega/2\pi$ is the estimated frequency of the signal, $t_0$ is the time origin and
\begin{equation}\label{eq:A}
A_q(\omega) = \frac{1}{N}\sum_{n=0}^{N-1}q_n \exp(-i\omega n\Delta t).
\end{equation}
is the discrete Fourier transform of $q$ evaluated at $\omega$.
Here, $q_n= q(t_n)$ is the measurement sample of the time trace at time $t_n = t_0 + n\Delta t$, $N$ is the number of samples entering the estimation and $\Delta t$ is the sampling interval.
The estimation of the amplitude and the phase relies on precise estimation of the frequency $\Omega$.
We estimate $\Omega$ by maximising \eqref{eq:A} with respect to $\omega$, i.e. $A(\Omega) = {\rm max}( A(\omega) )$.
The width of the function $A(\omega)$, and thereby our ability to localise the maximum, depends on the length of the time trace $q(t)$.
Therefore, we use a long time trace measured over $T_{\rm meas.} = 0.1\, \rm s$ and sampled at $625$ kilosamples/second to estimate $\Omega$.
Subsequently, we use that value of $\Omega$ and Equs. \eqref{eq:aML} and \eqref{eq:thetaML} to estimate the instantaneous amplitude and phase from short parts of that same time trace.
The short parts of the time trace contain $N=160$ samples, corresponding to an integration over $32$ particle oscillations. This constitutes a good compromise between sufficient data points for an accurate estimation of $R$ and $\phi$, and fast time resolution to resolve the dynamics of the energy and phase fluctuations.
Note that maximising \eqref{eq:A} allows us to estimate the frequency with much better accuracy than $1/T_{\rm meas.} $.

The absolute phase of a harmonic oscillator is a time delay with respect to some time reference.
Without such a time reference the absolute phase is arbitrary and has no meaning.
However, the relative phase between two oscillators is meaningful, because one oscillator serves as a time reference to determine the phase of the other oscillator with respect the first oscillator.
Formally, this is expressed as
\begin{equation}\label{eq:Delta_phi_a}
\Delta \phi = {\rm arg} \left[A_p\cdot \left[A_m^{\frac{\Omega_p}{\Omega_m}}\right]^*\right] = (\phi_p - \frac{\Omega_p}{\Omega_m}[\phi_m + k 2 \pi]),
\end{equation}
where $A_p$ and $A_m$ are the Fourier transforms of the two signals, respectively (c.f. \eqref{eq:A}), and k is an integer which takes into account that the 
phase is only determined up to modulo $2\pi$.
Note that the exponent $\Omega_p/\Omega_m$ takes care that \eqref{eq:Delta_phi_a} does not depend on $t_0$.
Without loss of generality, we set $\phi_m = 0$, i.e. we choose our time origin such that it coincides with a maximum of the signal with frequency $\Omega_m$. 
For the special case of a parametrically driven particle, which oscillates at half the frequency of the parametric modulation ($ \Omega_m = 2 \Omega_p$), we get $\Delta \phi =  \phi_p -k\pi$.
Therefore, the above method allows to estimate the relative phase between the particle oscillation and the parametric modulation up to a multiple of $\pi$.

\subsection{Parameter estimation}
\begin{figure}[tb]
 \centerline{
    \includegraphics[height=4.5cm]{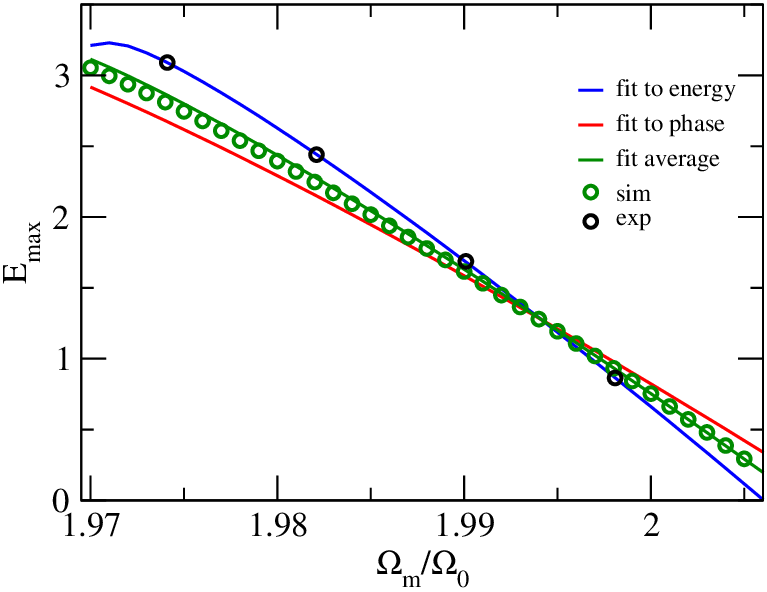}
    \includegraphics[height=4.5cm]{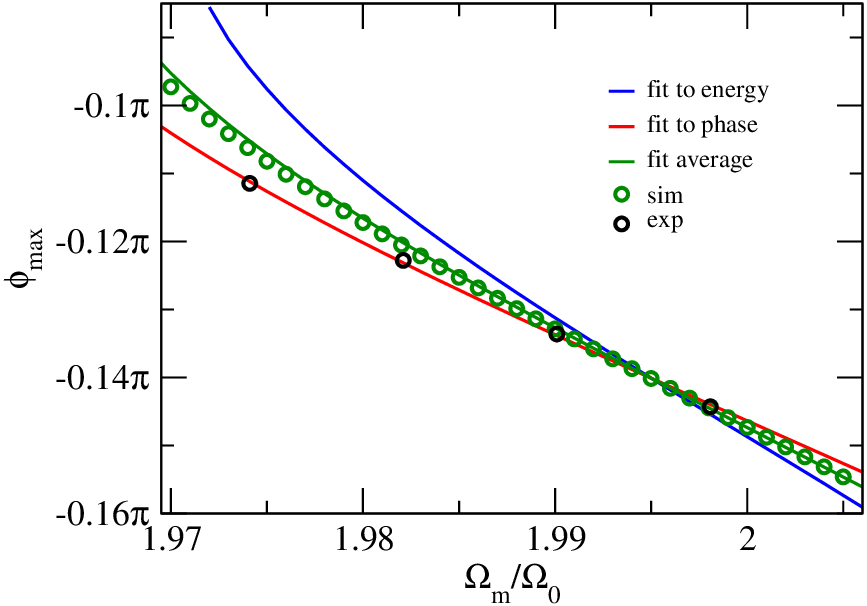}
    }
 \caption{
{\em Left:}
Most likely energy.
{\em Right:}
Most likely phase.
The black and green circles are the experimental data points and simulation results, respectively. The blue and red solid lines are the theoretical predictions for parameters obtained from independent fits to the energy and phase, respectively and the green solid line is the theoretical prediction for the averaged parameters.
 }
\label{fig:ExpMeanValues}
\end{figure}

We measure the distribution of the energy and phase for modulation at $\Omega_m/2\pi = 247, 248, 249$, and $250\rm\, kHz$.
Each distribution is obtained from $100$ time traces of $0.1\rm s$ duration.
Fig.~\ref{fig:ExpMeanValues} shows the maximum values of the energy and phase distributions shown in Fig.~\ref{fig:ExpDist} and a fit to secular perturbation theory \cite{Gieseler_PRL2014,LifshitzCross2008}.
While independent fits to the energy and phase, shown in blue and red, respectively, yield excellent agreement with the theoretical model, we cannot fit a set of parameters that would agree with both the energy and the phase.
Note that the phase fit includes a constant phase offset $\phi_0 = 50^\circ $ to account for the finite response time of the intensity modulator and delays in the electronics.
Averaging the results from the independent fits to energy and phase  yields $\xi = -5.4\pm 1.1\, \mu m^{-2}$, $\eta = 3.9\pm 1.3\,\mu m^{-2}$ and  $  \zeta = 16.1\pm 5.7 \times 10^{-3}$.
The theoretical curve for the parameters obtained by the energy and phase is shown in green together with numerical simulations using the parameters summarised in Table \ref{tab:ExperimentalParameter}.

The main uncertainty in the determination of the experimental parameters arises from the estimation of the particle mass and the resulting uncertainty in the voltage calibration and from parameter fluctuations, which we discuss in the next section.
As an independent measurement, we also measure the energy distribution without parametric modulation ($\zeta = 0$).
A fit of the energy distribution to Eq.~\eqref{eq:EnergyDist_FBon} yields  $\eta = 4.5\pm0.9\, \mu m^{-2}$, in good agreement with the previously determined value.

\subsection{Distributions}
\begin{figure}[tb]
 \centerline{
    \includegraphics[height=4.5cm]{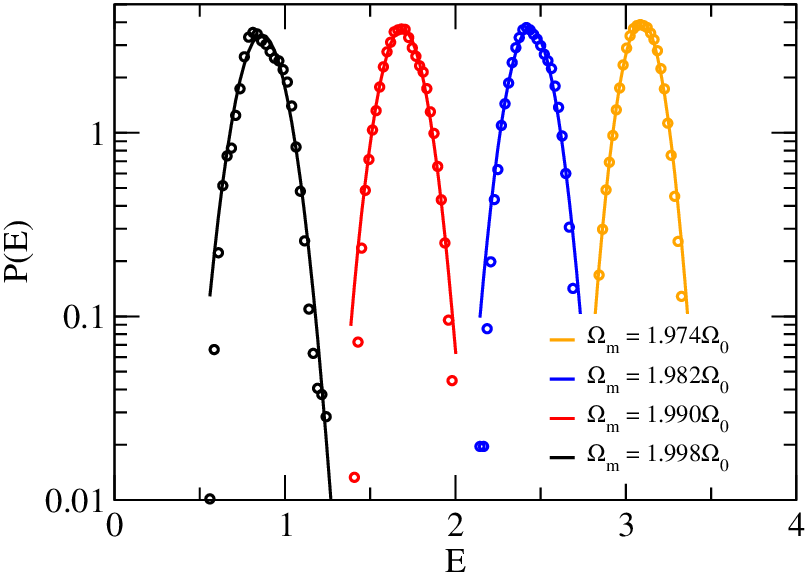}
    \includegraphics[height=4.5cm]{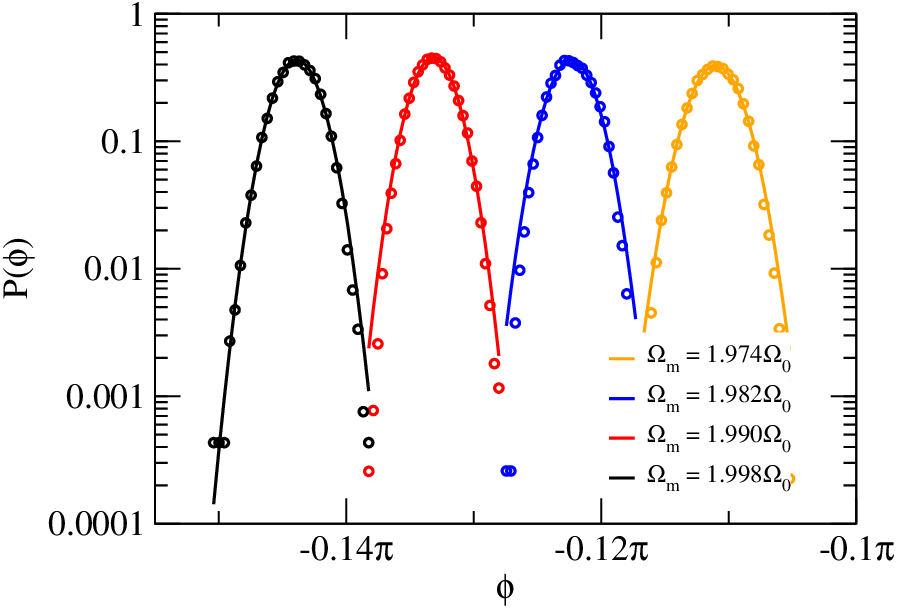}
    }
 \caption{
{\em Left:} Experimental distributions of energy
{\em Right:} Experimental distributions of phase.
The circles are experimental data points and the solid lines are Gaussian fits. The maxima correspond to the data points shown in Fig. \ref{fig:ExpMeanValues}.
 }
\label{fig:ExpDist}
\end{figure}

\begin{figure}[tb]
 \centerline{
    \includegraphics[height=4.5cm]{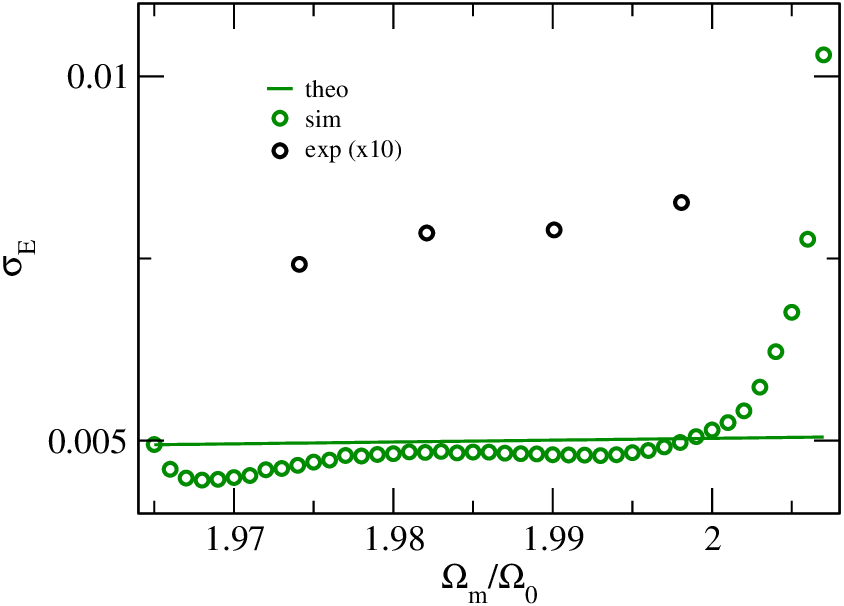}
    \includegraphics[height=4.5cm]{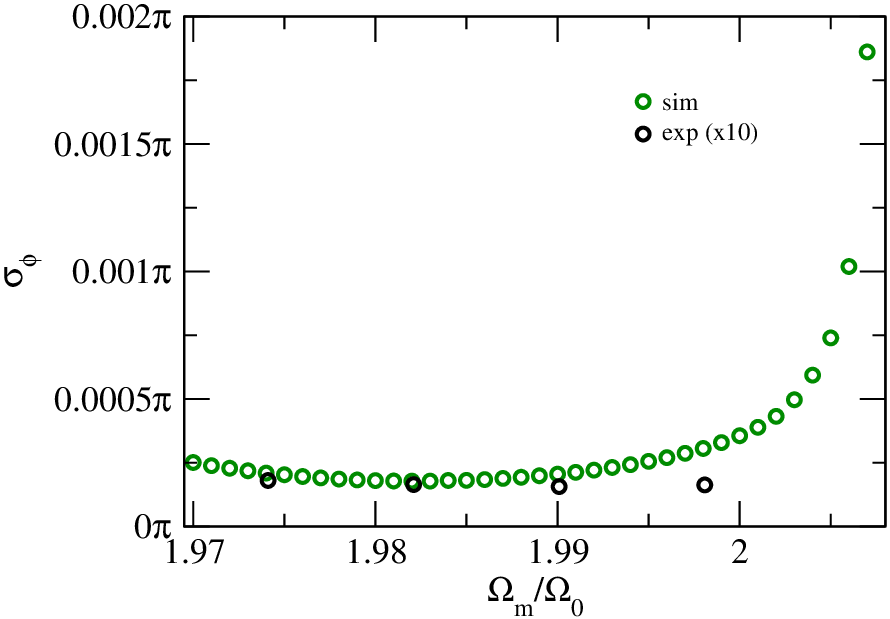}
    }
 \caption{
{\em Left:} Widths of energy distributions.
{\em Right:} Widths of phase distributions.
The black circles are experimental data points obtained form the Gaussian fits in Fig. \ref{fig:ExpDist}. The green circles are simulation results and the green solid line is the theoretical prediction Eqn. \eqref{eqn:EnergyVar}.
Note that the experimental values are significantly larger than the theoretical ones and are scaled by a factor of $0.1$ to fit them into the same plotting range.
 }
\label{fig:WidthDist}
\end{figure}

Fig.~\ref{fig:ExpDist} shows the experimental energy and phase distributions fitted with a Gaussian. 
As predicted by our theory and simulations, the distributions are Gaussian and their widths depend only weakly on the modulation frequency.
Fig.~\ref{fig:WidthDist} shows the widths of the distributions obtained from the Gaussian fits in Fig.~\ref{fig:ExpDist} and from numerical simulations.
For comparison, we also show the theoretical prediction according to Equ. \eqref{eqn:EnergyVar}.
The broadening of the distributions has two contributions, thermal motion and parameter fluctuations.

Thermal motion of the resonator, caused by residual air molecules, enters directly as a random white noise, which we considered in our theoretical model.
In addition, it enters indirectly through amplitude-phase conversion \cite{Kenig_PhysRevE_2012}.
The latter contribution has not been considered in our theoretical model but is naturally present in the numerical simulations.
Amplitude-phase conversion refers to the interdependence of energy and phase (c.f. \eqref{eqn:EnergyMean}).
Therefore, fluctuations in the phase cause fluctuations in the energy and vice versa.
This leads to a broadening of the distributions near the instability boundaries, where the deviation of the numerical simulation from our model is largest. 
Within this range, on the other hand, this interplay manifests itself as sidebands in the power spectral density of the particle position \cite{Gieseler_PRL2014}.

In addition to Brownian motion, parameter fluctuations broaden the experimental distributions \cite{Villanueva_PRL_2013}.
The experimental parameters fluctuate due to laser intensity and polarisation fluctuations and also due to the nonlinear coupling with the other two degrees of freedom, which were not considered in our model \cite{Gieseler_NATPHYS,Gieseler_PRL2014}.
Noise in the feedback electronics and modulator gives rise to further broadening. 
In general, broadening due to fluctuating parameters dominates broadening due to Brownian motion. As a consequence, the measured width of the energy and phase distributions $\sigma_E = 78\pm 3 \times 10^{-3}$ $k_BT$ and $\sigma_\phi = 1.7 \pm 0.1 \times 10^{-3} \pi$, respectively, are approximately one order of magnitude larger than the theoretical values $5.1  \times 10^{-3}$ $k_BT$ and $0.15 \times 10^{-3} \pi$, averaged over the range of detunings of the experimental data.
To identify the noise sources responsible for the deviation from theory one can deliberately introduce noise and systematically study its effect on the measurement outcome.

\section{Conclusion}
\label{sec_conclusion}

We have developed a stochastic model for the dynamics of the energy of a nonlinear nanomechanical oscillator subject to parametric modulation and nonlinear damping. Under these conditions the oscillator attains a non-equilibrium steady state. Our model allows us to predict the energy distribution of the steady state. The steady state distribution is intimately related to fluctuation theorems, which describe the statistical properties of the system for transitions between different states \cite{NAT_NANOTECH}. Consequently, our work opens the door to test these fluctuation theorems in different scenarios. 

We confirmed the validity of the model by extensive numerical simulations and found excellent agreement with our theory.
In addition, we performed experiments with a levitated nanoparticle.
While the measured mean energy and phase are in close agreement with the numerical simulations, their distributions are broadened due to parameter fluctuations that are not accounted for in the theory and are subject to further investigation. Besides quantifying additional noise sources experimentally, future work includes the development of a more generalised model including a stochastic model for the phase and incorporating other white and non-white noise sources, resulting from fluctuating parameters \cite{Moehlis_2014}.

\section*{Acknowledgments}
This research was supported ERC-QMES (no. 338763), and the Austrian Science Fund (FWF) within the SFB ViCoM (grant F41). C.M. is supported by a uni:docs-fellowship of the University of Vienna. 


\end{document}